\documentclass[a4paper,
              ]{jpconf}
 {\usepackage[utf8]{inputenc}}           
\usepackage[british,english,american]{babel}
\usepackage[final]{pdfpages}
\usepackage{multirow}
\usepackage{ragged2e}
\usepackage[caption=false]{subfig}
 
\graphicspath{{figures/}}

\begin{document}

\title{Amplitude enhancement of the self-modulated plasma wakefields}

\author{Y. Li\textsuperscript{1,2}, G. Xia\textsuperscript{1,2}, K. V. Lotov\textsuperscript{3,4}, A. P. Sosedkin\textsuperscript{3,4}, Y. Zhao\textsuperscript{1}, S. J. Gessner\textsuperscript{5}}
\address{\textsuperscript{1}University of Manchester, Manchester, UK \\ 
\textsuperscript{2}Cockcroft Institute, Daresbury, UK\\
\textsuperscript{3}Budker Institute of Nuclear Physics, Novosibirsk, Russia\\
\textsuperscript{4}Novosibirsk State University, Novosibirsk, Russia\\
\textsuperscript{5}CERN, Geneva, Switzerland}

\ead{yangmei.li@manchester.ac.uk}	

%
\begin{abstract}
Seeded Self-modulation (SSM) has been demonstrated to transform a long proton bunch into many equidistant micro-bunches (e.g., the AWAKE case), which then resonantly excite strong wakefields. However, the wakefields in a uniform plasma suffer from a quick amplitude drop after reaching the peak. This is caused by a significant decrease of the wake phase velocity during self-modulation. A large number of protons slip out of focusing and decelerating regions and get lost, and thus cannot contribute to the wakefield growth.  Previously suggested solutions incorporate a sharp or a linear plasma longitudinal density increase which can compensate the backward phase shift and therefore enhance the wakefields. In this paper, we propose a new plasma density profile, which can further boost the wakefield amplitude by 30$\%$. More importantly, almost 24$\%$ of protons initially located along one plasma period survive in a micro-bunch after modulation. The underlying physics is discussed.

\end{abstract}

\section{INTRODUCTION}

Proton bunches with huge energies have been proposed to drive electrons to TeV energies in a single plasma stage \cite{item:1}. Successful exploitation of the effect assumes the proton bunch to be as short as hundreds of $\mu$m while currently producible ones are dozens of cm long. This is orders of magnitude longer than the plasma wavelengths of interest, which makes strong wake excitation virtually impossible. Nevertheless, in 2010 Kumar \textit{et al.} \cite{item:2} proposed that with proper seeding the long bunch will see an efficient growth of the transverse wakefields generated by itself and be modulated by the periodically focusing and defocusing forces. This self-modulation chops the long proton bunch into many equidistant micro-bunches which are one plasma wavelength apart. Strong wakefields are excited accordingly. In addition, the seeded self-modulation (SSM) dominates over the non-axisymmetric hosing instability, which otherwise would destroy the bunch.    

The first proof-of-principle experiment AWAKE studies accelerating electrons with a long self-modulated proton bunch \cite{item:3}. In the experiment, a laser pulse used for gas ionization co-propagates with the long proton bunch, starting in its middle. In this case, only the rear half of the bunch with a sharp-cutting front edge propagates in the plasma, and the self-modulation is strongly seeded. However, the wake phase velocity decreases significantly during the development of SSM \cite{item:4,item:5}. The large backward phase shift forces a large number of protons into the defocusing regions, which results in dramatic proton loss and wake amplitude decrease after reaching the peak \cite{item:6,item:7,item:8}. 

There are studies of longitudinal plasma profiling \cite{item:6,item:7,item:8}, where a density increase helps to reduce the local plasma wavelength so that the phase shift can be compensated. The plasma wakefield amplitude has been promoted, but suboptimally \cite{item:8}. In this paper, we accommodate the need for more complicated plasma density profiles and propose a new plasma shape which can further boost the wakefield.

\section{SIMULATIONS}
  \subsection{Baseline}

\begin{figure}[!b]
 \begin{minipage}[l]{22pc}
  \subfloat{
   \includegraphics[width=22pc]{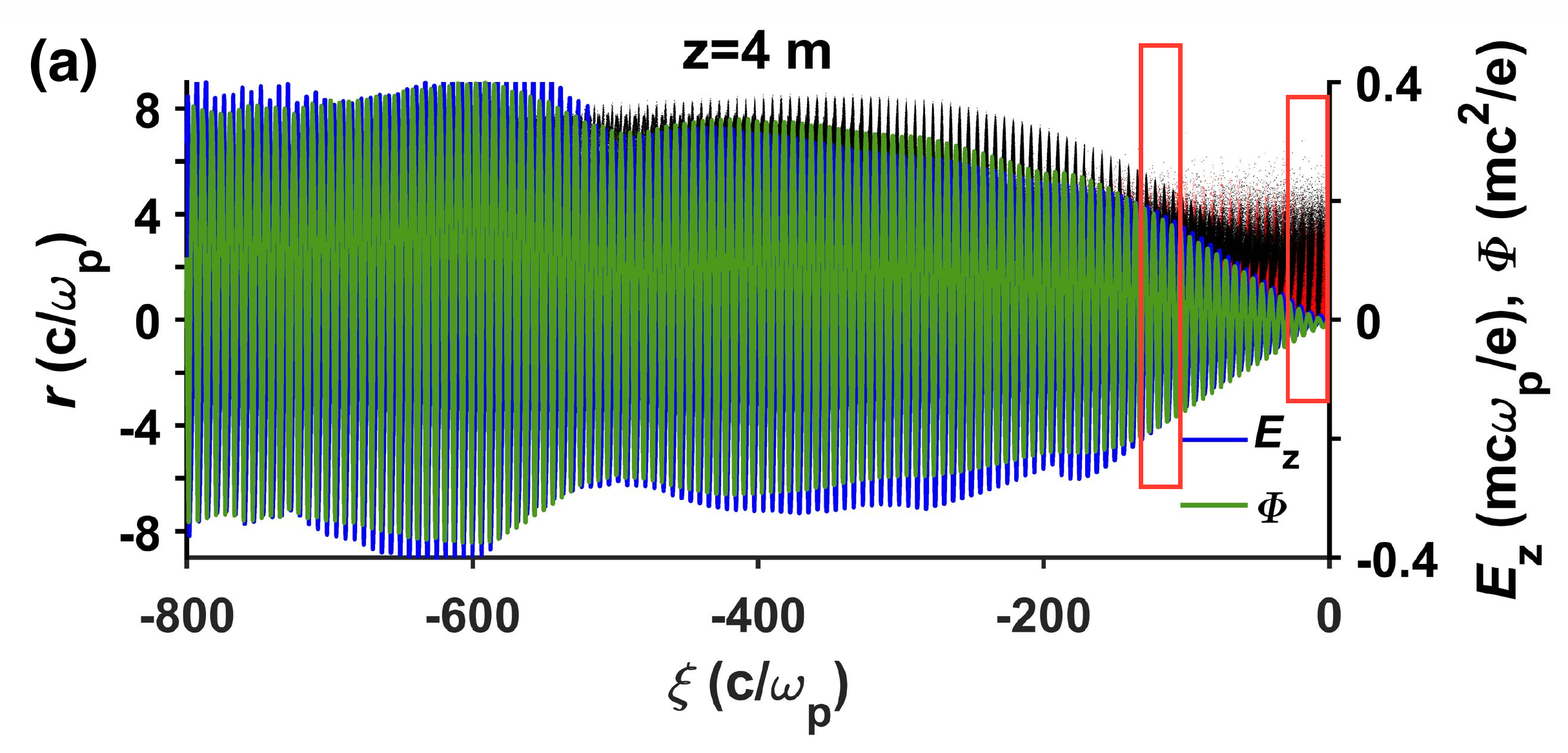}
  }
  \newline
  \subfloat{
   \includegraphics[width=11pc]{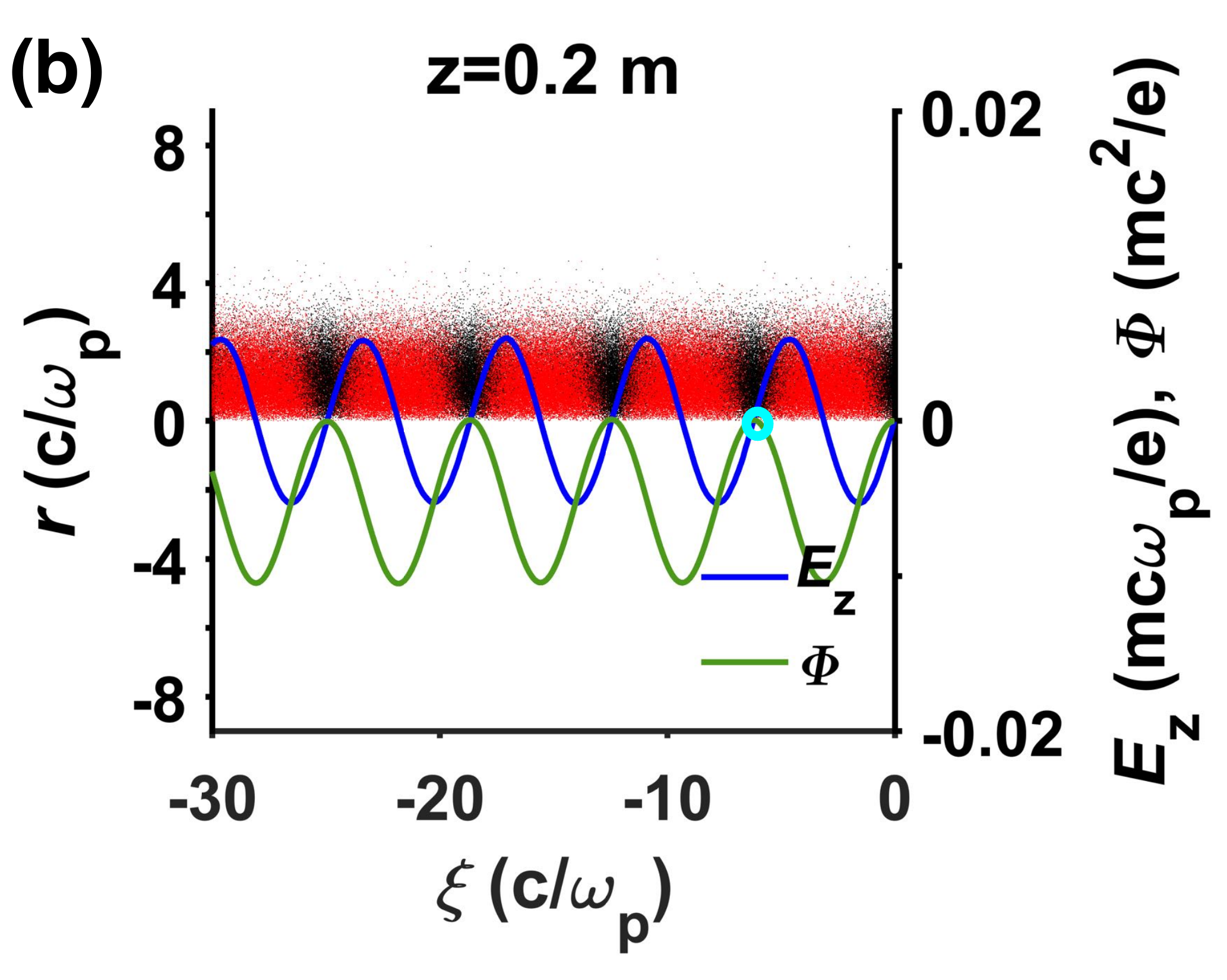}
  }
  \subfloat{
   \includegraphics[width=11pc]{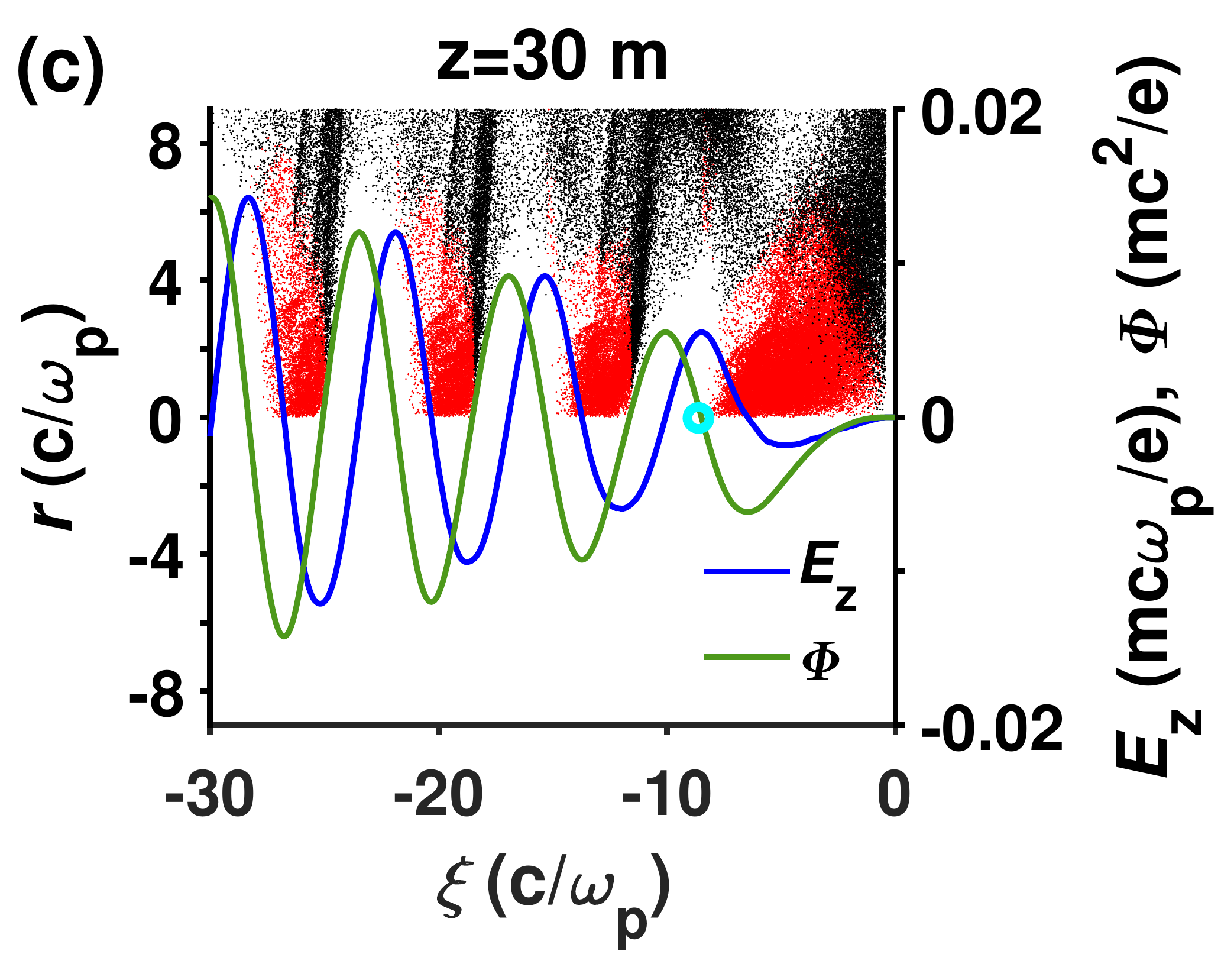}
  }
  \newline
  \subfloat{
   \includegraphics[width=10pc]{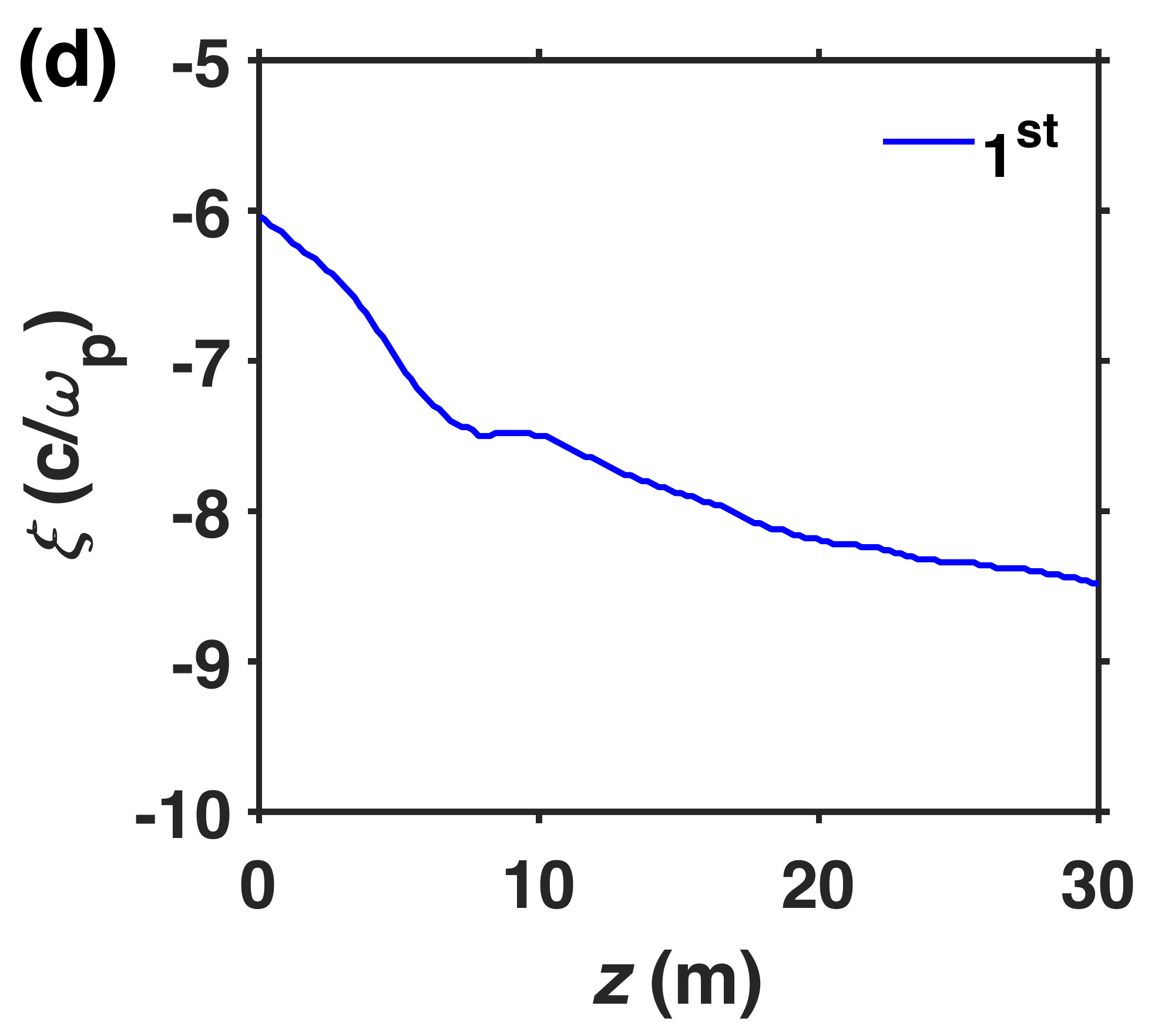}
  }
  \subfloat{
   \includegraphics[width=12pc]{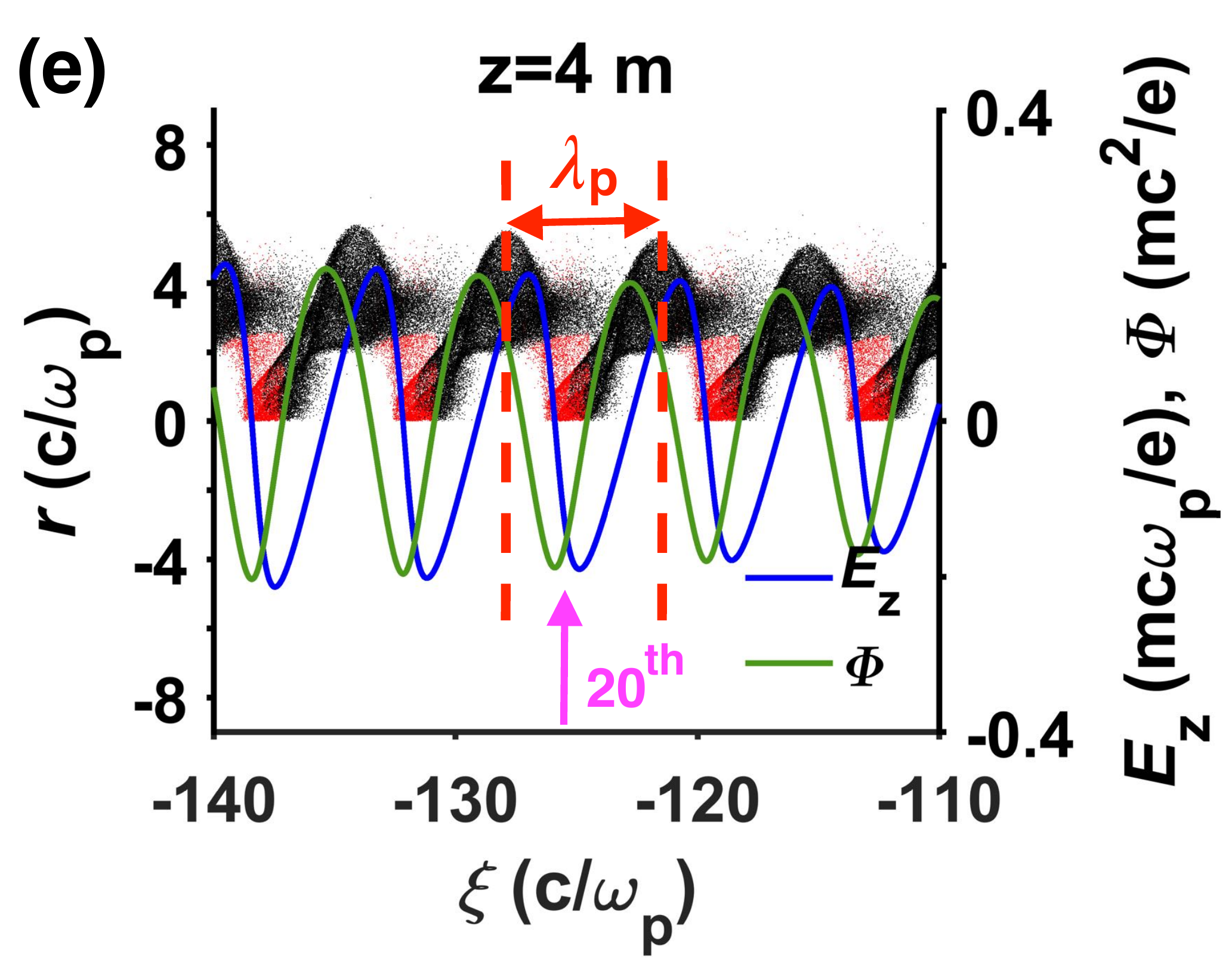}
  }
\end{minipage}\hfill
\begin{minipage}[c]{14pc}
\caption{\label{label}Bunch distribution in real space (points), on-axis longitudinal electric field $E_z$ (blue curve), and wakefield potential $\Phi$ (green curve) for the whole beam at $z=4$\,m (a), for the bunch head ($\xi\in~$[-30, 0]) at $z=0.2$\,m (b) and $z=30$\,m (c), and for the front part ($\xi\in~$[-140, -110]) at $z=4$\,m (e). Red points denote the protons that are confined by potential wells, while black points represent protons capable of escaping. (d) Longitudinal position $\xi=z-ct$ of the first potential zero [two positions therein are marked by cyan circles in (b) and (c)] versus the propagation distance~$z$.}
 \label{fig:baseline}
\end{minipage}
\end{figure}

With a 2D axisymmetric quasi-static particle-in-cell code LCODE \cite{item:9}, we first simulate a twice shorter proton beam as compared to the baseline AWAKE scenario \cite{item:3}. This beam length is envisaged for AWAKE upgrades (Table~\ref{table1}). The beam propagates in $z$-direction (to the right in Fig.\,\ref{fig:baseline}a). Only the rear half that propagates in plasma is simulated, starting with an abrupt cut, so as to mimic SSM seeding by the ionization front. With periodic focusing and defocusing transverse forces excited (i.e., negative and positive wakefield potentials shown by green lines), some protons are confined (marked as red points) and others diverge (marked black). This way the long bunch is chopped into many micro-bunches. 

\begin{table}[!t]
   \caption{Proton and Plasma Parameters}
   \begin{center}
   \begin{tabular}{lll}
\br
       \textbf{Parameters} & \textbf{Values}                      &\textbf{Units} \\
\mr
           Bunch population, $N_p$                         & 3~$\times$~$10^{11}$              &               \\ 
           Bunch energy, $W_0$                        & 400                                           &GeV        \\ 
           Energy spread, $\delta W/W$              & 0.7$\%$                                           &                  \\ 
           RMS length, $\sigma_z$                 & 6                                              &cm              \\
           RMS radius, $\sigma_r$                 & 200                                           &$\mu$m      \\
           Normalized emittance, $\epsilon_n$              & 3.5                                              &$\mu$m                 \\
           Plasma density, $n_0$                         & 7~$\times$~$10^{14}$                &cm$^{-3}$  \\
\br
   \end{tabular}
   \end{center}
   \label{table1}
\end{table}

Ideally, the micro-bunches are approximately $\lambda_p = 2 \pi c/\omega_p$ apart, where $\omega_p = \sqrt{4 \pi n_0 e^2/m_e}$ is the plasma frequency, and other notations are common. Each bunch occupies $\lambda_p/4$, corresponding to both decelerating and focusing region. However, before the micro-bunching, the wake potential near the bunch head has larger regions with negative values than with positive ones. From Figs.\,\ref{fig:baseline}b and~\ref{fig:baseline}c, we see protons along $\xi\in[-2\pi,\,0]$  are focused and located in the negative potential region. Therefore, the first micro-bunch occupies almost a full wave period after self-modulation and shifts the wave phase back by almost 2.5$c/\omega_p$ with respect to the initial perturbation (Fig.\,\ref{fig:baseline}d). Note that the nonlinear elongation of the wake period due to increasing wake amplitude along the bunch is negligible here \cite{item:10}. As SSM develops, plenty of protons drop into defocusing regions and get lost because of the significant backward phase shift. Take the $20\textsuperscript{th}$ micro-bunch for instance (Fig.\,\ref{fig:baseline}e), we see only 2$\%$ of protons survived (Fig.\,\ref{fig:survival}). Without doubt, the wakefield drops drastically with such huge proton loss (Fig.\,\ref{fig:wake}a).

\subsection{Plasma Profiling}

\begin{figure}[!b]
\begin{minipage}[c]{17pc}
\includegraphics[width=17pc]{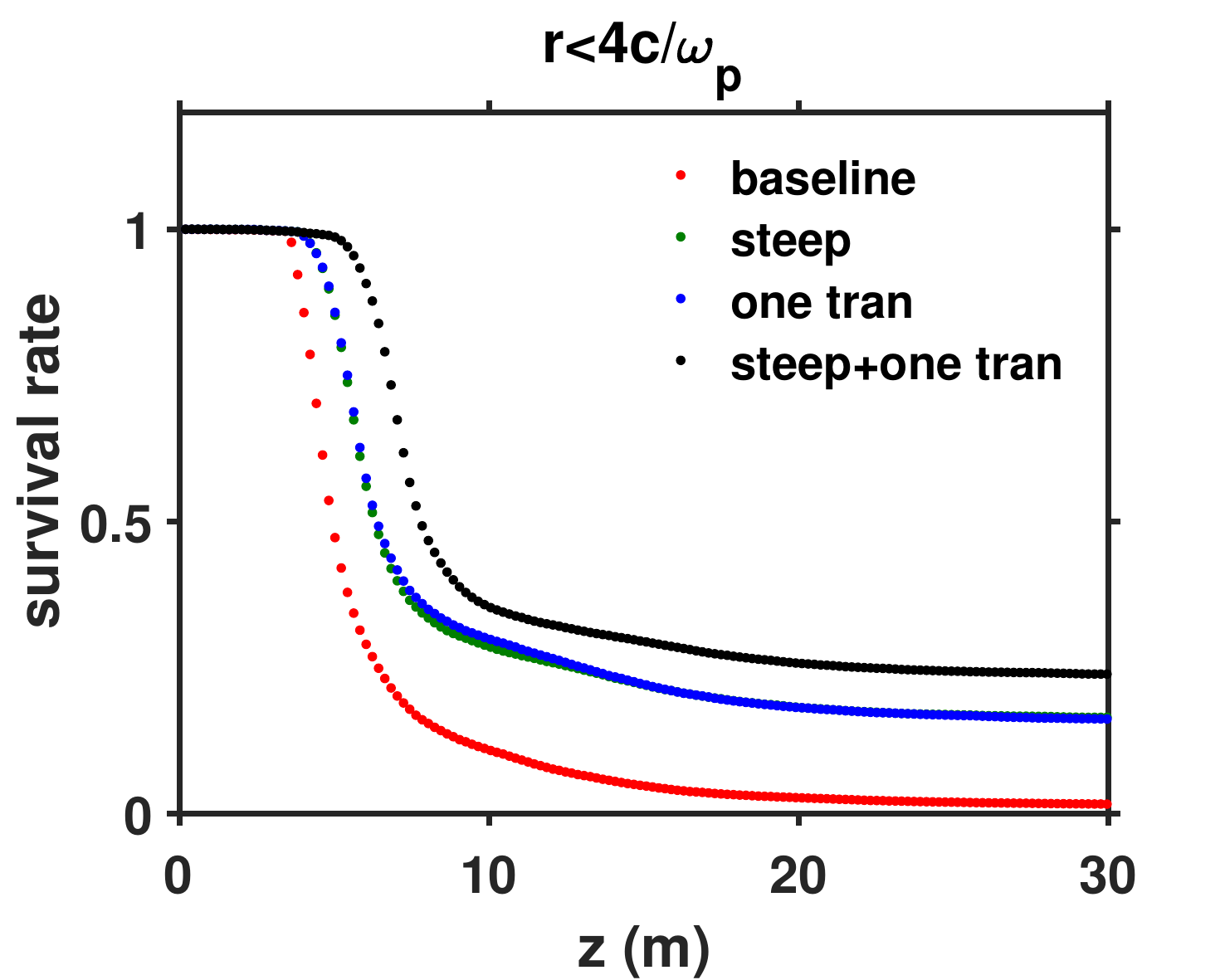}
\end{minipage}\hfill
\begin{minipage}[c]{20pc}
\caption{\label{label}Survival rates of protons initially located in the interval of the length $\lambda_p$ near the 20th bunch for all discussed cases. A proton is counted as survived if its radial position is smaller than 4$c/\omega_p$.}
 \label{fig:survival}
\end{minipage} 
\end{figure}

\begin{figure}[!t]
\begin{minipage}[c]{22pc}
  \subfloat{
   \centering
   \includegraphics[width=11pc]{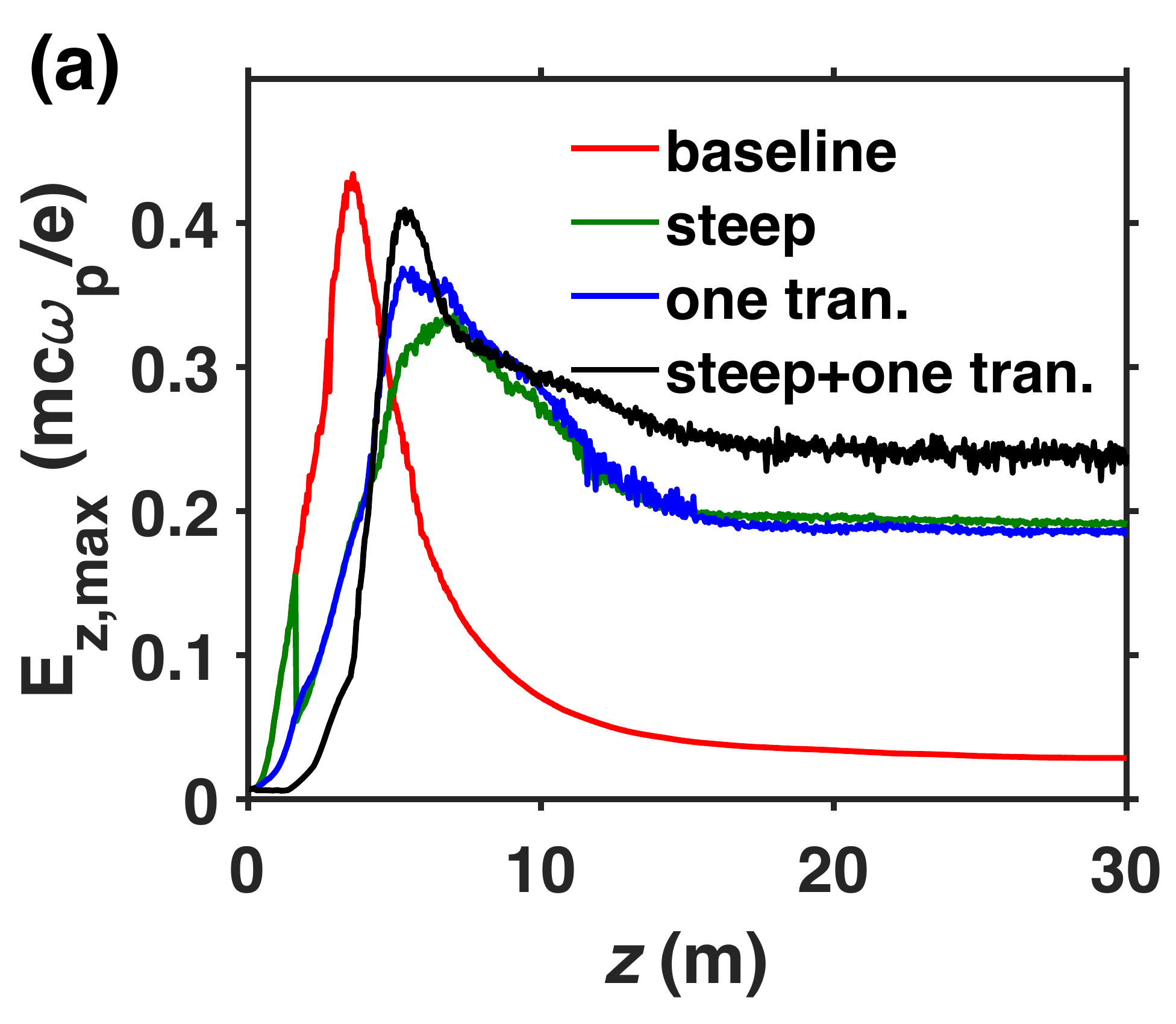}
  }
  \subfloat{
   \centering
   \includegraphics[width=11pc]{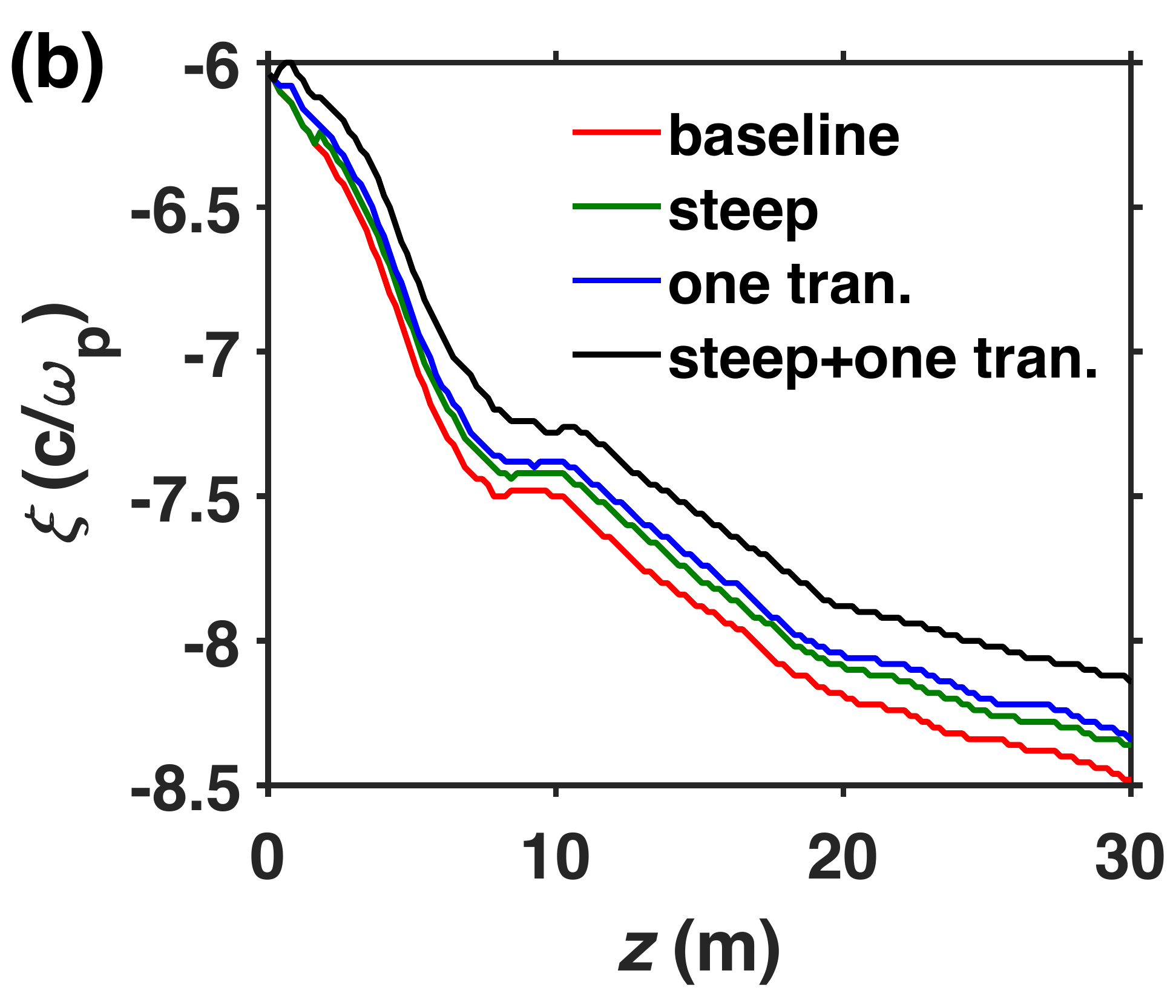}
  }
  \newline
  \subfloat{
   \centering
   \includegraphics[width=11pc]{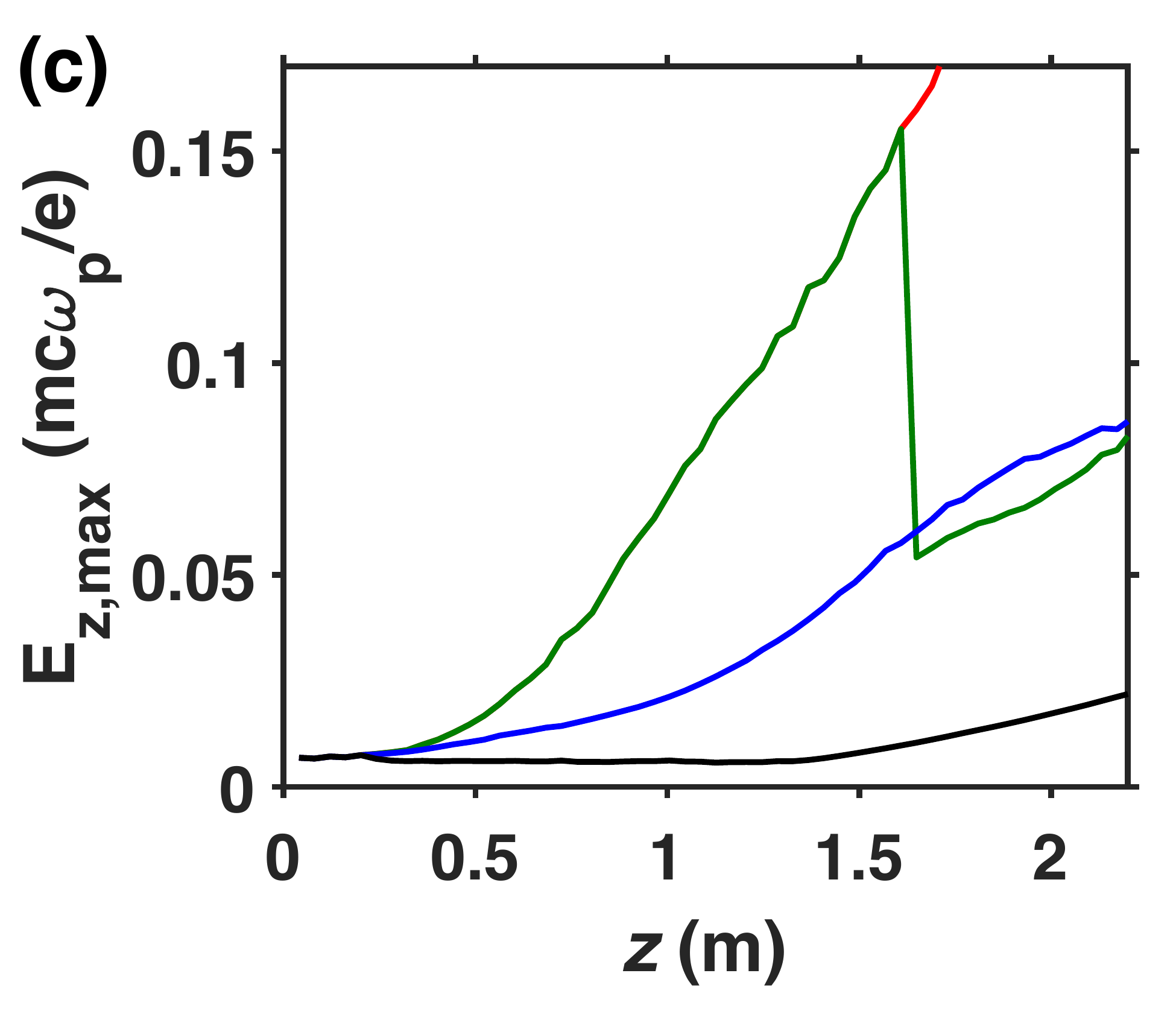}
  }
  \subfloat{
   \centering
   \includegraphics[width=11pc]{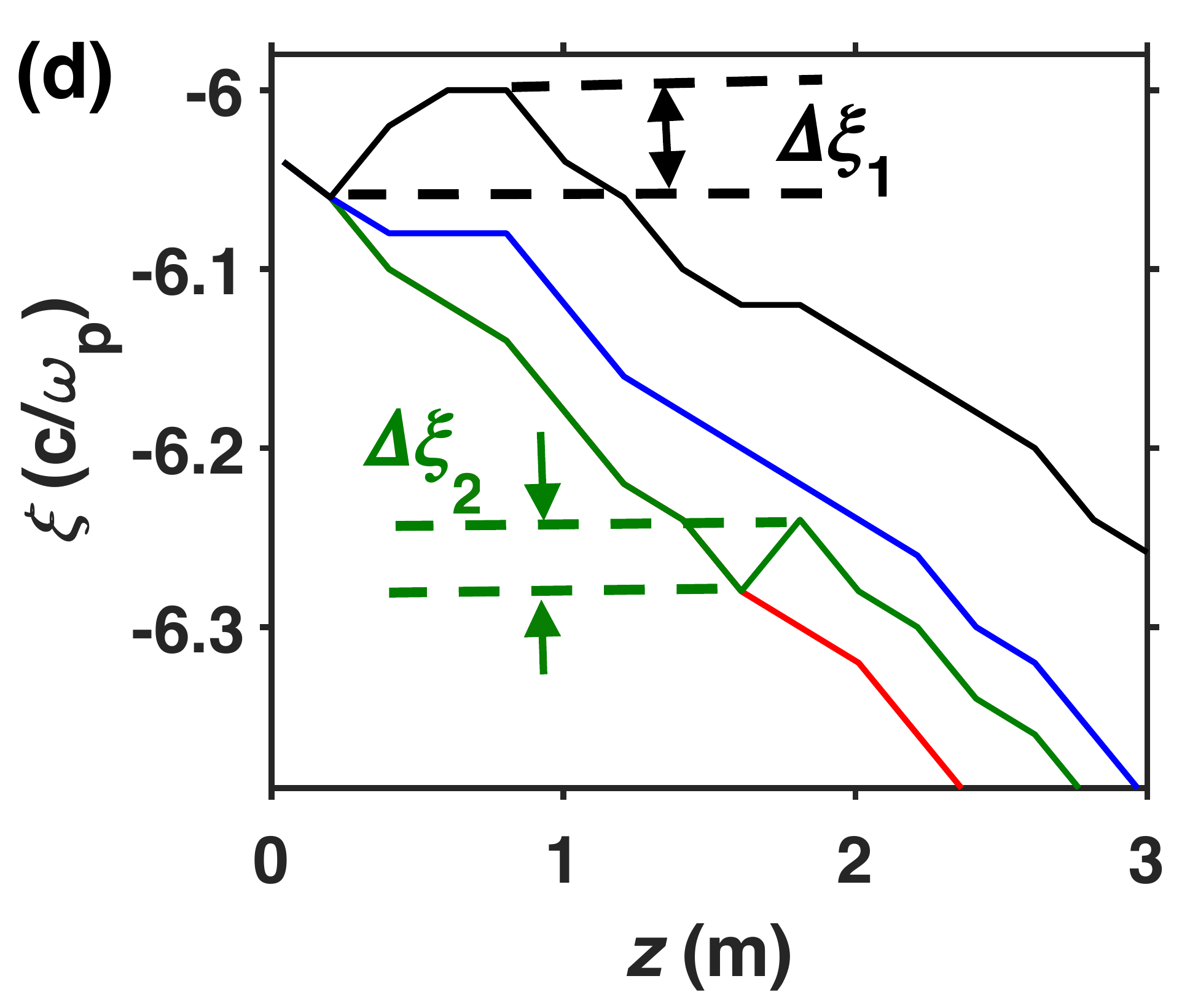}
  }
\end{minipage}\hfill
\begin{minipage}[c]{14pc}
 \caption{\label{label}Maximum wakefield amplitudes (a) and longitudinal positions of the first potential zero (b) versus the propagation distance for all discussed cases and their close-ups at the initial distance (c) and (d). The curves of the same colour in top and bottom plots represent the same case.}
 \label{fig:wake}
\end{minipage}
\end{figure}

\begin{figure*}[!t]
 \begin{center}
  \subfloat{
   \centering
   \includegraphics[width=9pc]{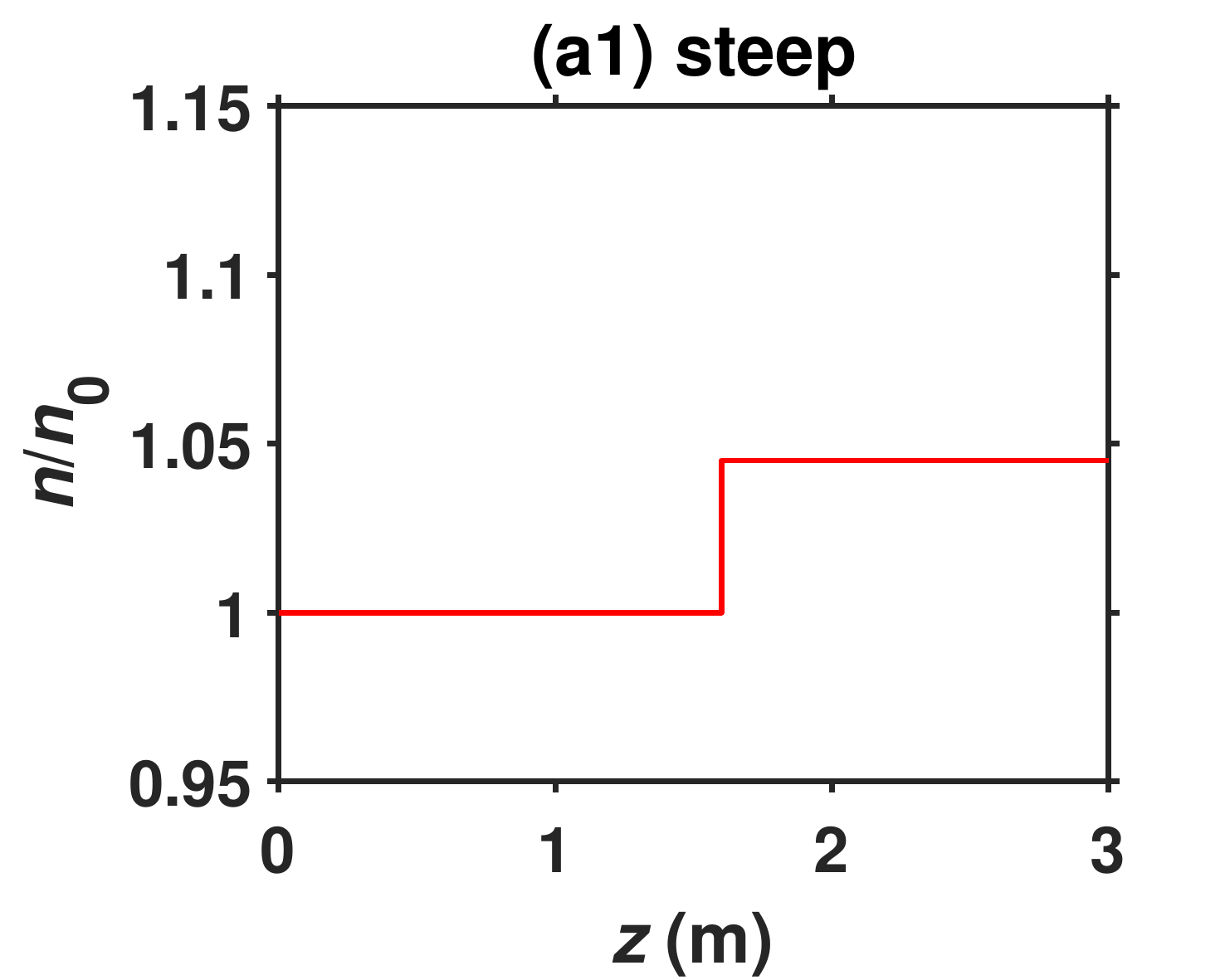}
  }
  \subfloat{
   \centering
   \includegraphics[width=9pc]{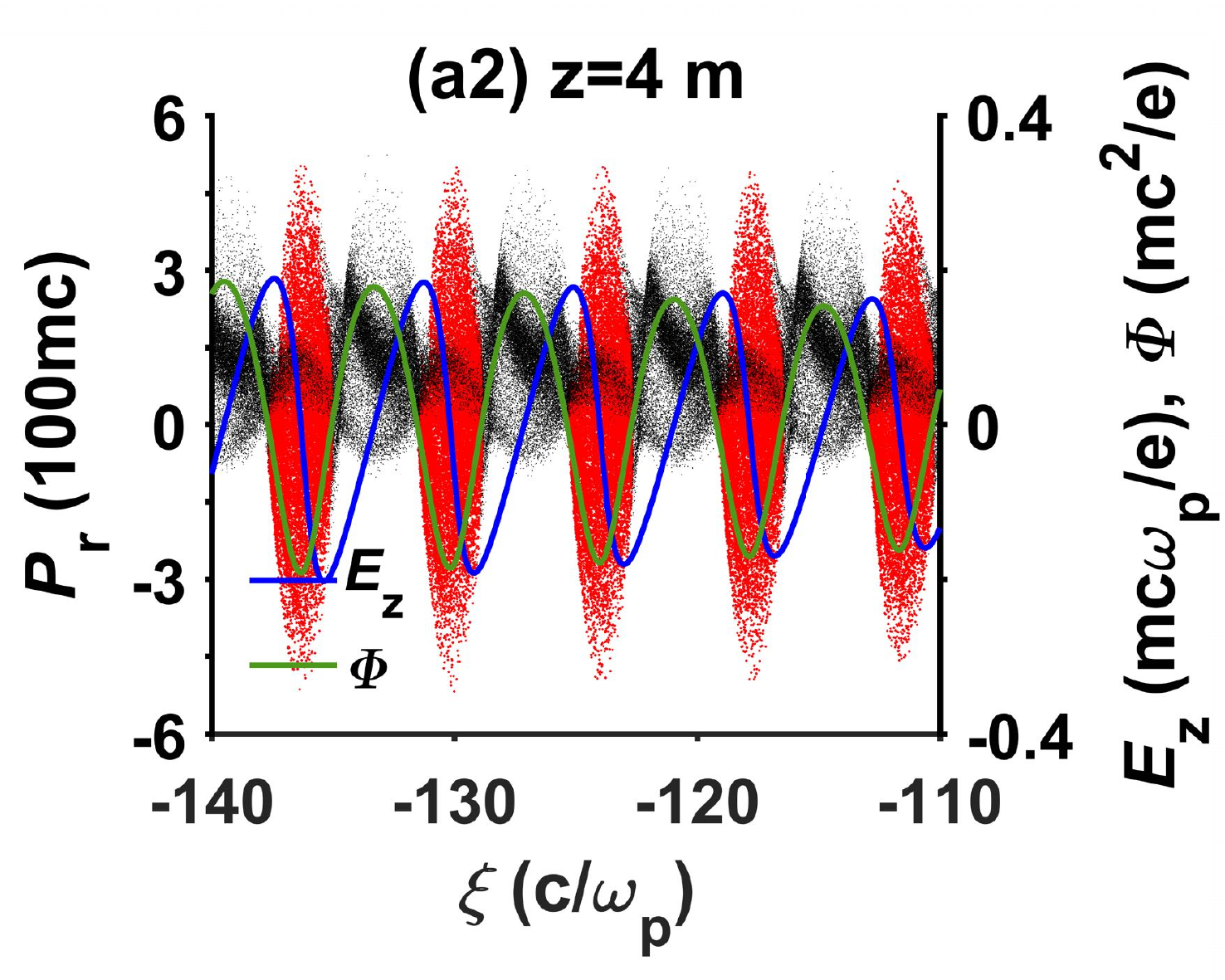}
  }
    \subfloat{
   \centering
   \includegraphics[width=9pc]{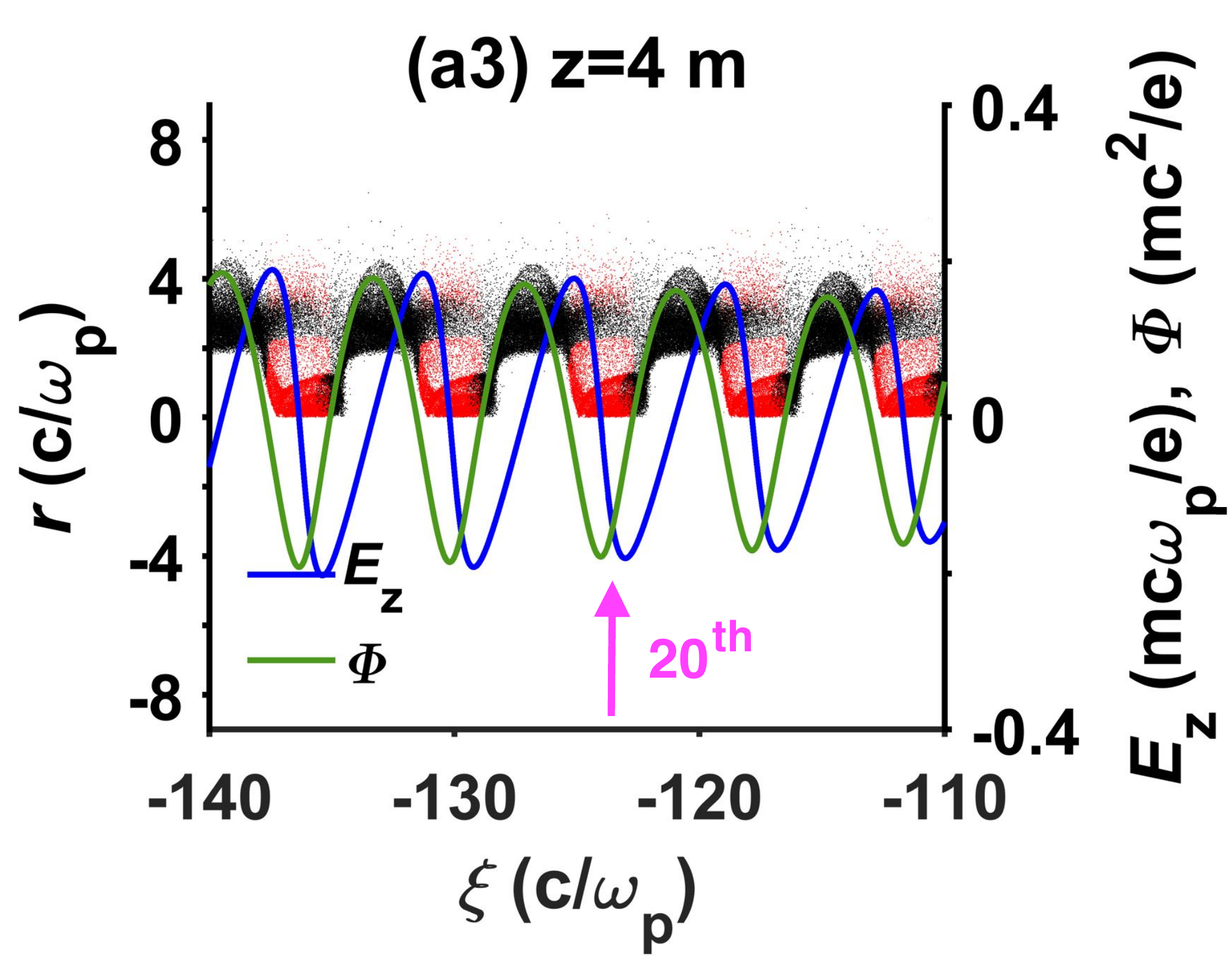}
  }
  \subfloat{
   \centering
   \includegraphics[width=9pc]{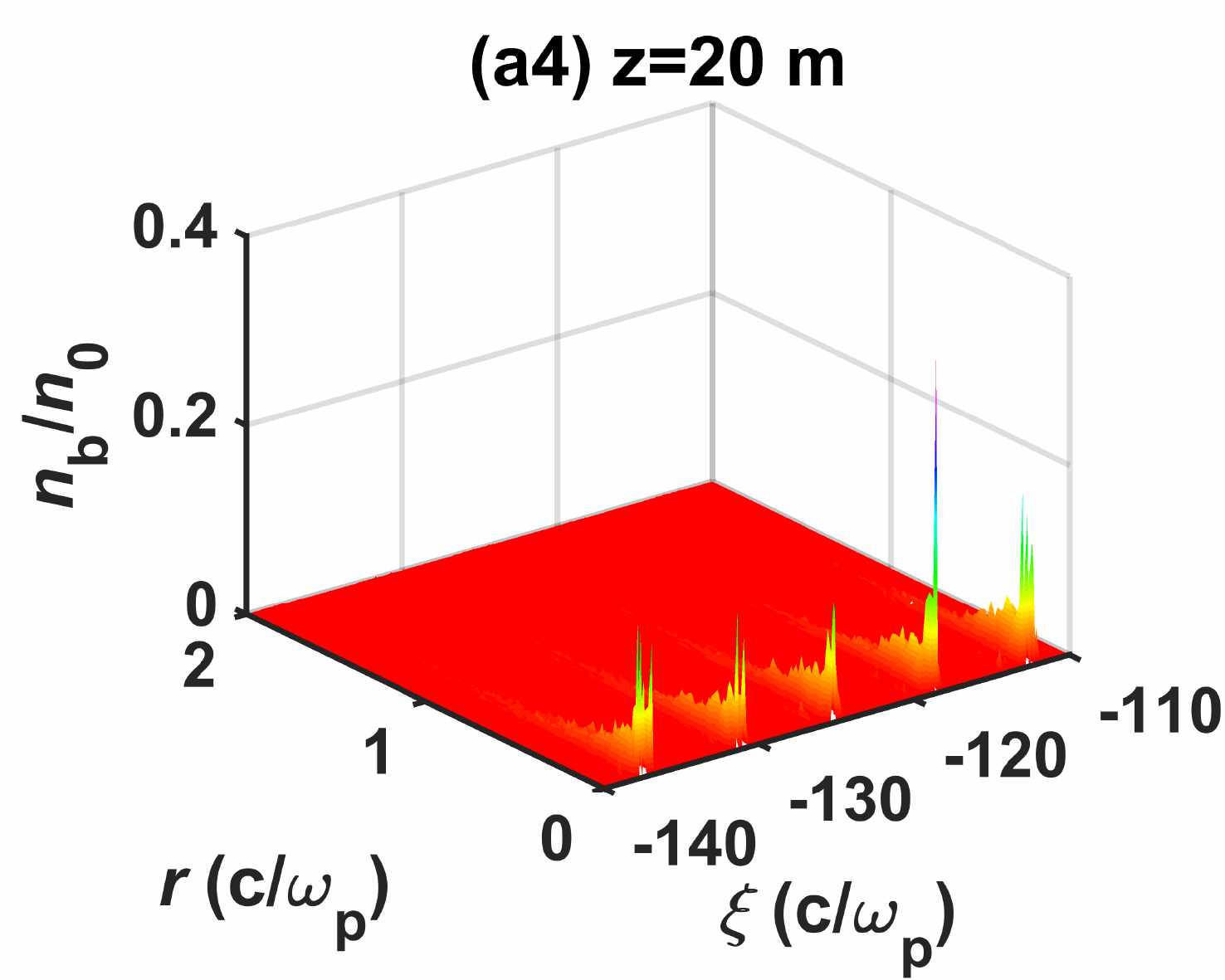}
  }
  \newline
  \subfloat{
   \centering
   \includegraphics[width=9pc]{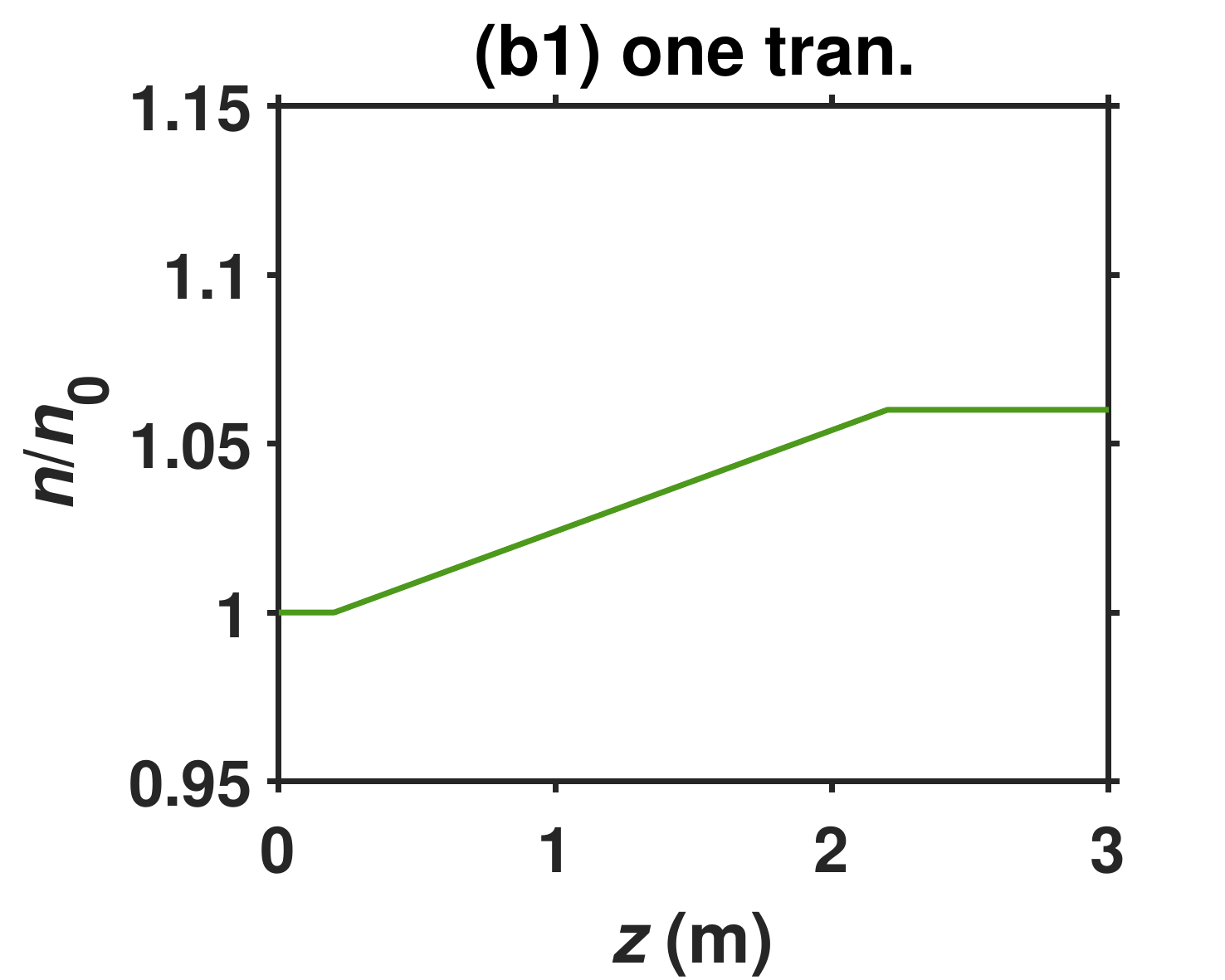}
  }
  \subfloat{
   \centering
   \includegraphics[width=9pc]{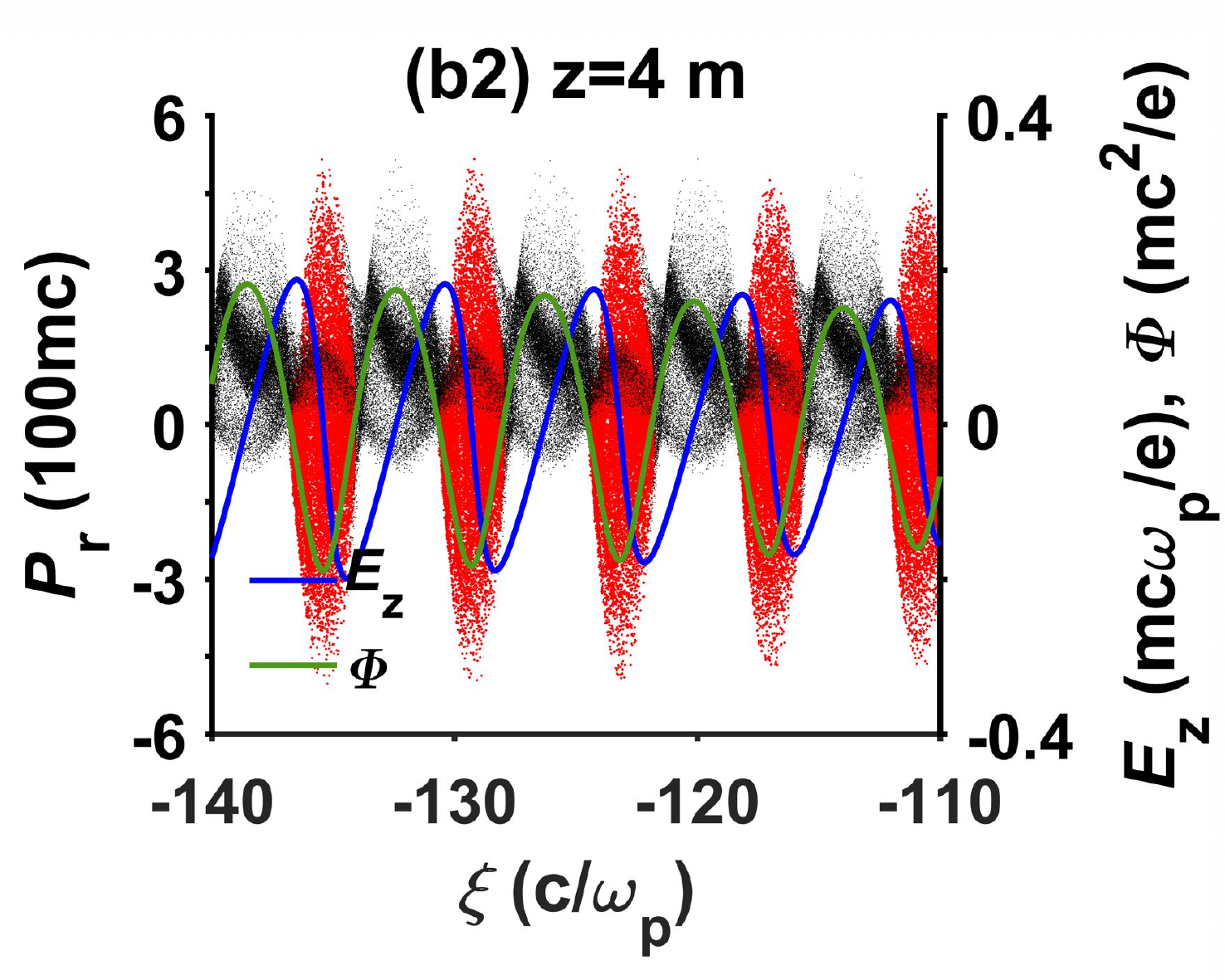}
  }
    \subfloat{
   \centering
   \includegraphics[width=9pc]{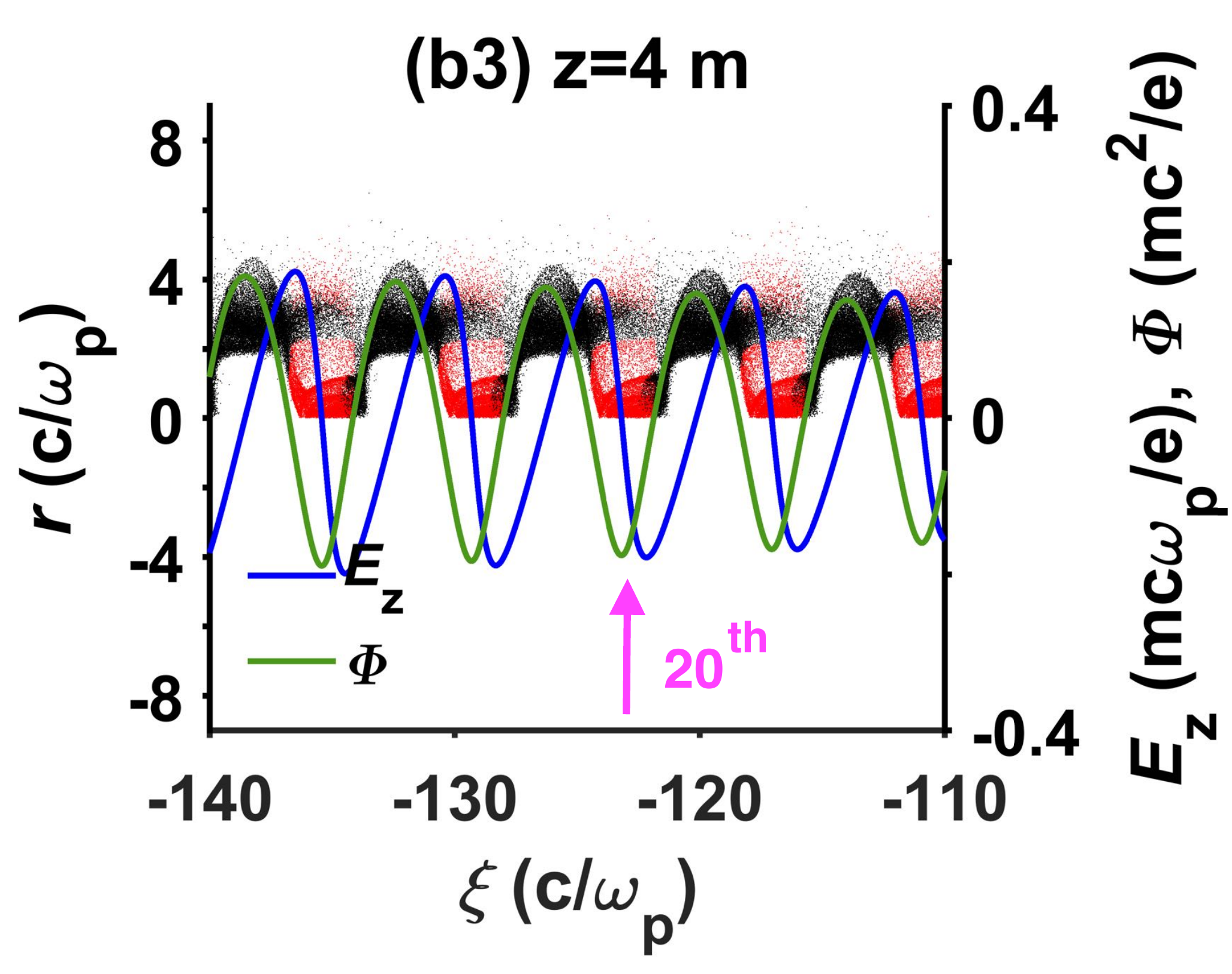}
  }
  \subfloat{
   \centering
   \includegraphics[width=9pc]{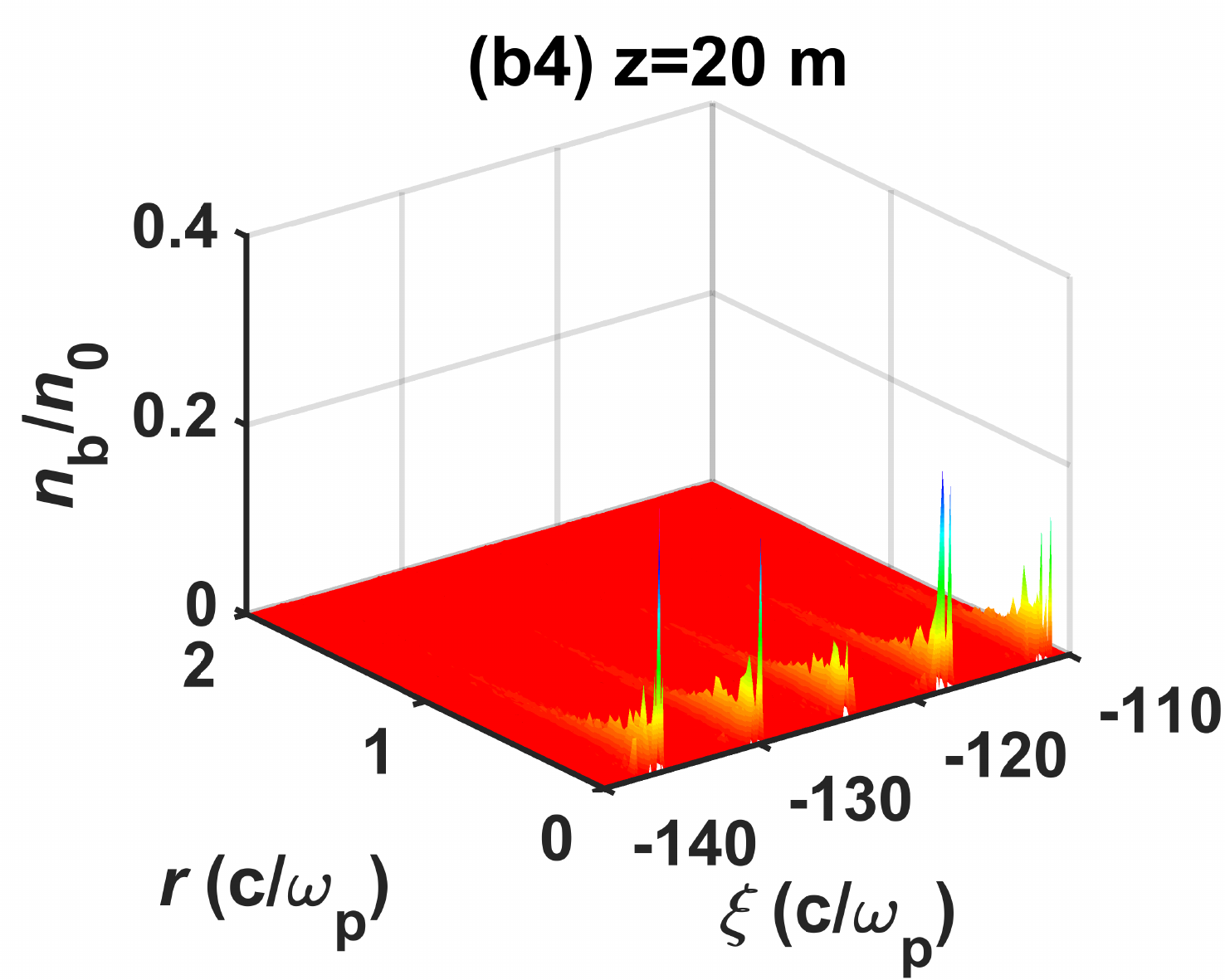}
  }
    \newline
  \subfloat{
   \centering
   \includegraphics[width=9pc]{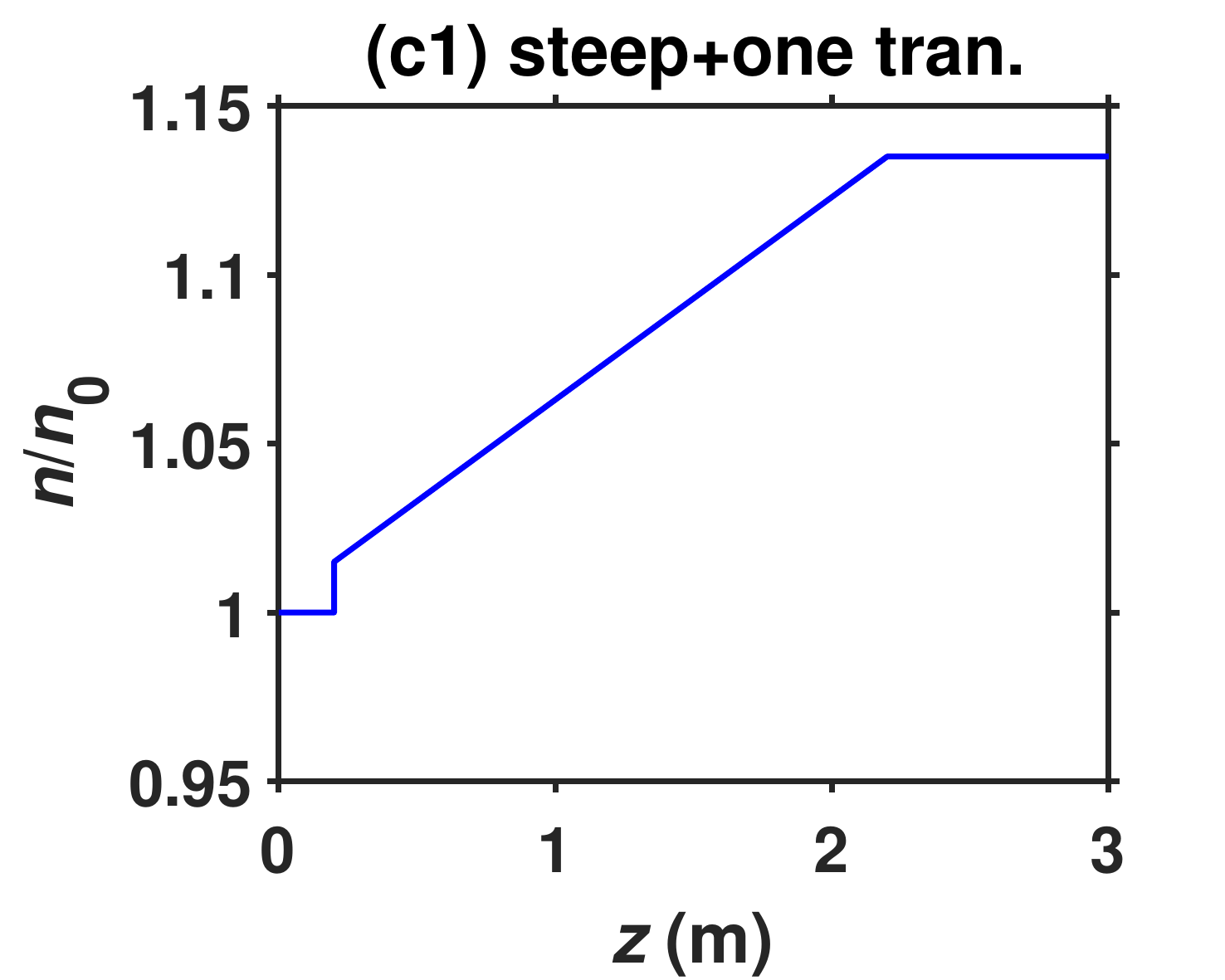}
  }
  \subfloat{
   \centering
   \includegraphics[width=9pc]{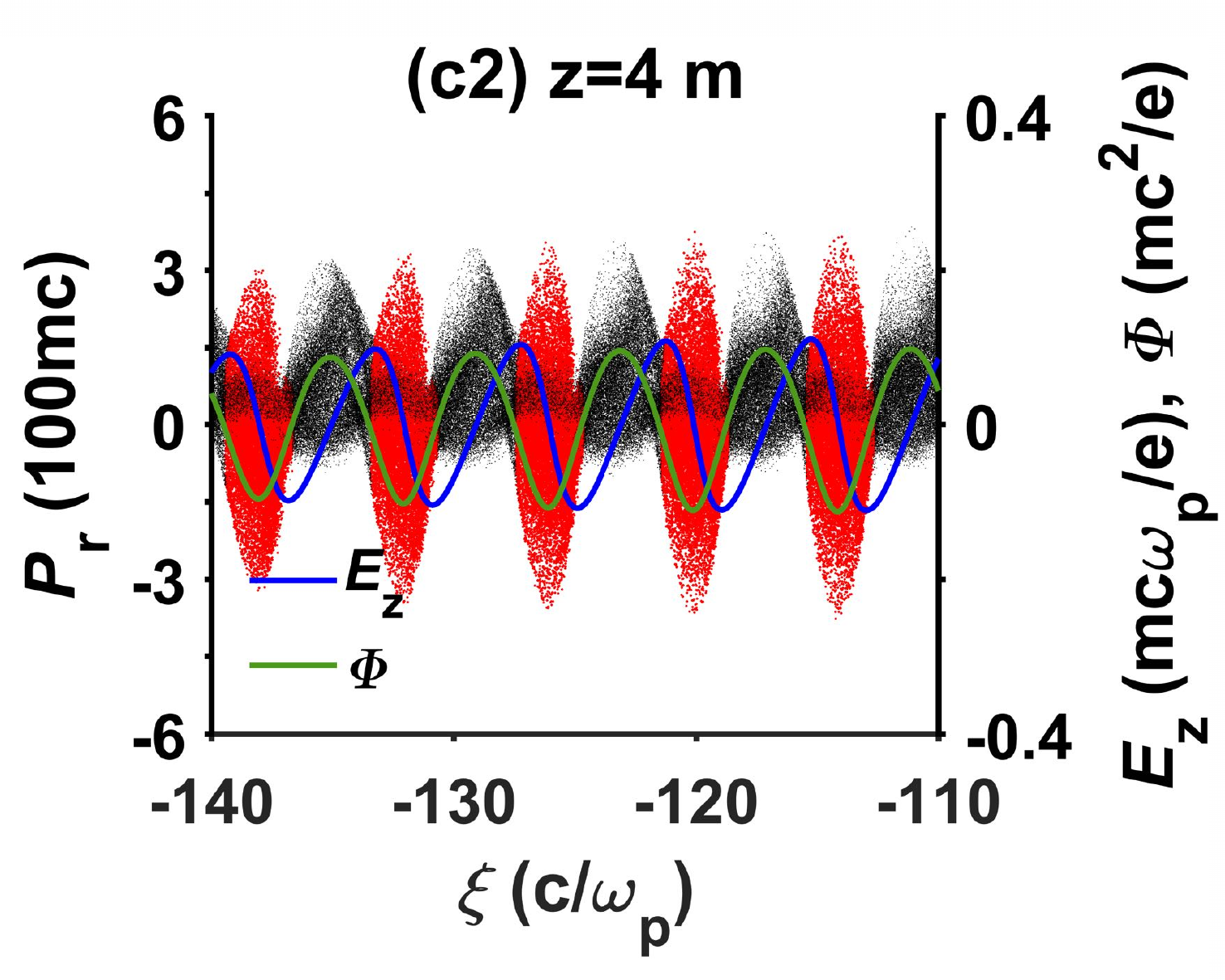}
  }
    \subfloat{
   \centering
   \includegraphics[width=9pc]{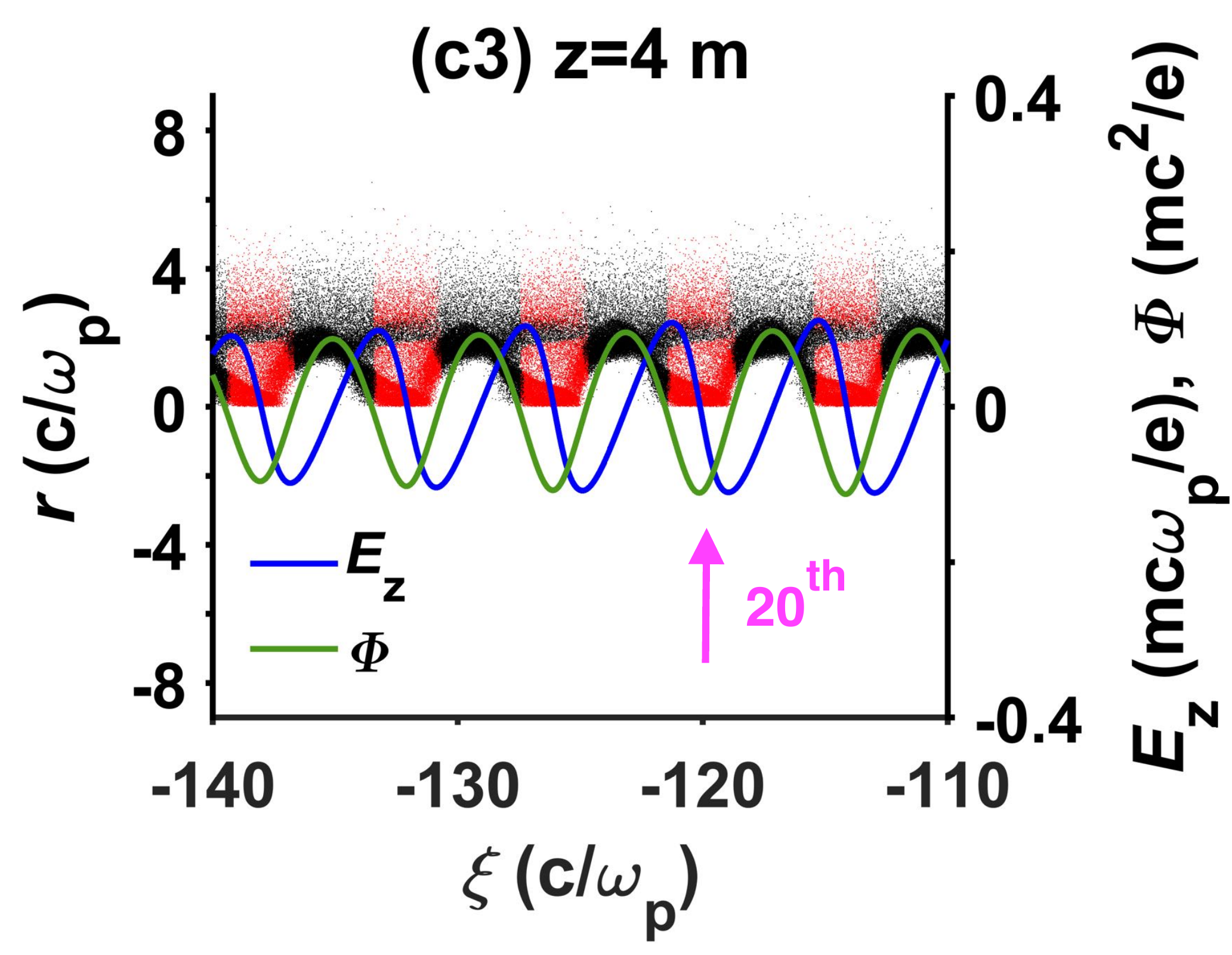}
  }
  \subfloat{
   \centering
   \includegraphics[width=9pc]{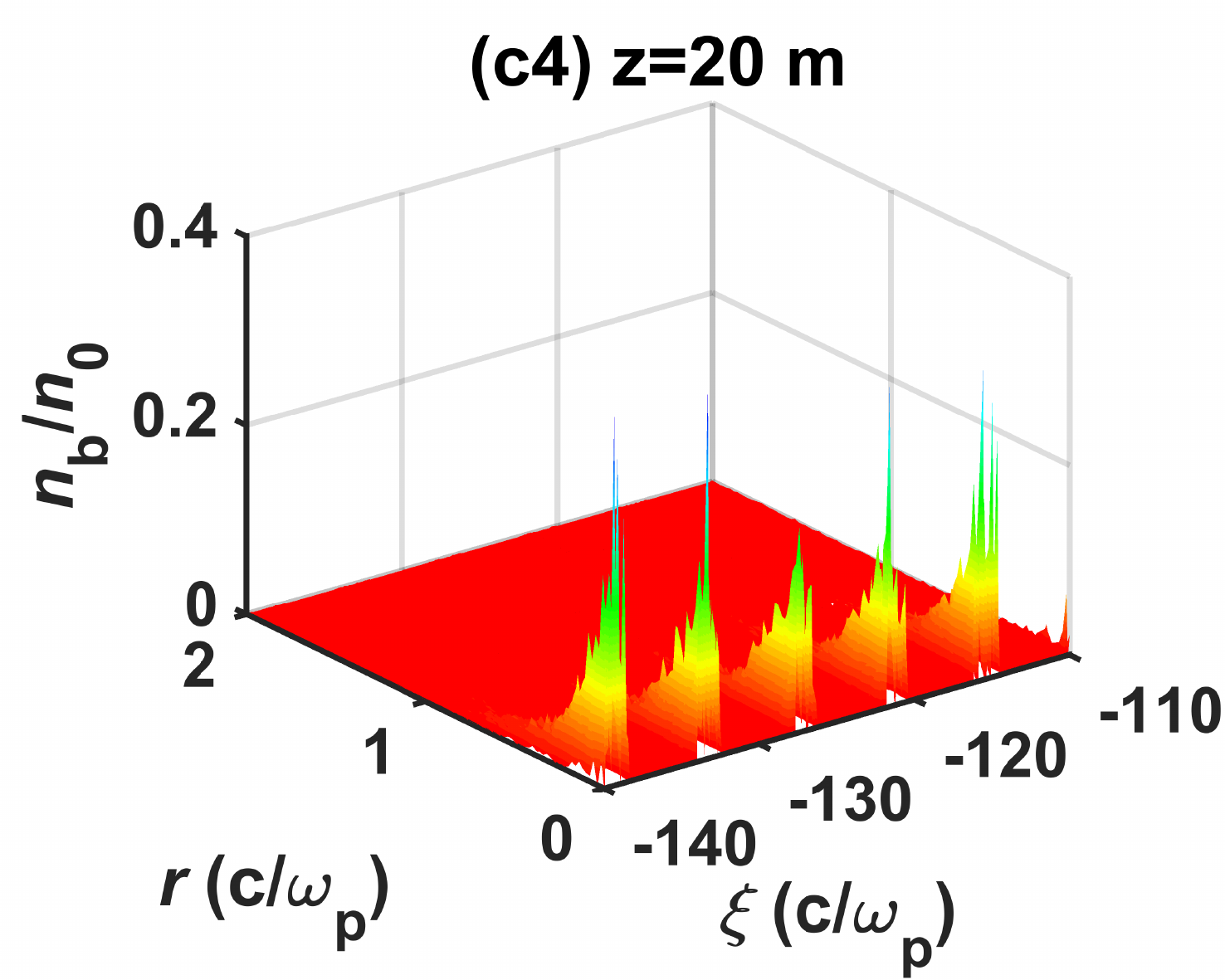}
  }
 \end{center}
 \caption{Plasma density profiles (1st column), proton radial momenta $P_r$ (2nd column) and positions $r$ (3rd column) at $z=4$\,m plotted together with on-axis electric field $E_z$ and wakefield potential $\Phi$, and micro-bunch density $n_b (r,\xi)$ at $z=20$\,m (4th column) for three considered cases. Proton colouring is the same as in Fig.\,\ref{fig:baseline}.}
 \label{fig:taper}
\end{figure*}

Since the drastic phase shift is the main issue hindering strong wake excitation based on SSM, researchers have proposed two typical types of longitudinal plasma density profiles: density step-up [noted as case 1, Fig.\,\ref{fig:taper}(a1)] and linear density transition [noted as case 2, Fig.\,\ref{fig:taper}(b1)]. Increase of the plasma density reduces the local plasma wavelength and thus brings forward the back shifted phase. With the same proton beam parameters given in Table 1 and the same initial plasma density, we first sort out the optimum plasma density profile for each case. "Optimum" here corresponds to the highest established wake amplitude that we measure at $z=30$\,m. For the case 1, the relative plasma density increases steeply by 4.5$\%$ at $z=1.6$\,m. The optimum linear transition corresponds to 6\% density increase in the interval from $z=0.2$\,m to $z=2.2$\,m. Fig.\,\ref{fig:wake}a confirms a substantial enhancement of the wake amplitudes by these two plasma profiles.

Here we propose a third plasma density profile. It features an early and steep density increase immediately followed by a linear transition. The optimum set-up is to increase the density steeply by 1.5$\%$ after 0.2\,m and then linearly until 13.5$\%$ in 2\,m. It resembles a simple combination of the first two cases but is capable of further enhancing the wake amplitude by about 30$\%$ (Fig.\,\ref{fig:wake}a).

The underlying reasons for the wakefield enhancement are twofold. First of all, introducing a small and sharp increase in the early stage can correct the phase shift more effectively. Fig.\,\ref{fig:wake}d demonstrates that the initial phase correction ($\Delta\xi_1$) in our proposed case (black curve) is much larger than the one ($\Delta\xi_2$) produced by the later steep density increase of a larger (6\%) value (case 1, green curve). Likewise, the linear density increase, although acting the same early (from $z=0.2$\,m), contributes less to the phase correction (case 2, blue curve). Another reason is, in our proposed case, the wakefield stays low during a long time (Fig.\,\ref{fig:wake}c). As a result, SSM development slows down, and protons experience smaller transverse forces in the early stage, thus gaining smaller transverse momenta [Fig.\,\ref{fig:taper}(c2)] than in other two cases [Figs.\,\ref{fig:taper}(a2) and \ref{fig:taper}(b2)]. Consequently, fewer protons have enough transverse momentum to escape from the potential well during micro-bunching. This is especially important for protons near the boundaries between positive and negative potentials, where the potential well is pretty shallow. We see a larger proton concentration near the axis and within the focusing and decelerating region [Fig.\,\ref{fig:taper}(c3)], which is favorable for wake excitation. As for the other two cases, protons located in the vicinity of zero potentials escape easily [marked black in Figs.\,\ref{fig:taper}(a3) and \ref{fig:taper}(b3)].  This is further confirmed by the micro-bunch density in Fig.\,\ref{fig:taper}(c4) being observably higher than in the other two cases [Figs.\,\ref{fig:taper}(a4) and \ref{fig:taper}(b4)].

Figure \ref{fig:survival} gives a quantitative measure of the proton loss in the $20\textsuperscript{th}$ micro-bunch for the discussed cases. In our proposed case, 24$\%$ of protons survive, which approaches the ideal value of 25$\%$ (quarter-period). For the first two cases, the survival rate is 16$\%$. Also protons in these cases get lost quicker. 

\section{CONCLUSIONS}
In this paper, we propose and investigate a slightly more complicated plasma profile than the usual steep or linear density increase. The main idea is to slow down the SSM in the early stage, so that protons gain smaller transverse momenta while being bunched and escape less from the potential wells, especially near micro-bunch boundaries. More protons are therefore kept within the favorable focusing and decelerating  region of the wave. Simulations illustrate that this new profile enables further boost of the wakefield amplitude by 30$\%$, and 24$\%$ of the initial beam charge remains in the micro-bunches.  

%
%

\section*{References}
\bibliography{iopart-num}

\end{document}